\documentclass[a4paper]{article}

\usepackage[english]{babel}
\usepackage[utf8x]{inputenc}
\usepackage[OT2,T1]{fontenc}
\usepackage{amssymb}

\usepackage[a4paper,top=3cm,bottom=2cm,left=3cm,right=3cm,marginparwidth=1.75cm]{geometry}

\usepackage{amsmath}
\usepackage{graphicx}
\usepackage[colorinlistoftodos]{todonotes}
\usepackage[colorlinks=true, allcolors=blue]{hyperref}

\title{Efficient Implementation of Baker–Campbell–Hausdorff Formula}
\author{Cupjin Huang}

\begin{document}
\maketitle

\begin{abstract}
This short paper presents an efficient implementation of Baker–Campbell–Hausdorff formula for calculating the logarithm of product of two possibly non-commutative Lie group elements using only Lie algebra terms.
\end{abstract}

\section{Introduction}

Given a Lie group $G$ and its Lie algebra $\mathfrak{g}$, there is an exponential map
$$\exp:\mathfrak{g}\rightarrow G.$$
In a small neighborhood of the identity element $I\in G$, $\exp$ is a smooth bijection and has an inverse map $\log:G\rightarrow \mathfrak{g}$. 

It is sometimes very useful to compute the logarithm of a product of two elements in the Lie group near the identity, i.e.\ $Z=\log(\exp{X}\exp{Y})$. In the case that $G$ is commutative, we can solve $Z$ exactly as $X+Y$; however, difficulty arises when $G$ is non-commutative. Our goal is to approximately compute $Z$ up to a given order $N$, which will be defined later.

\section{Dynkin's Explicit Expression for BCH formula}
Due to Eugene Dynkin, the explicit combinatorial expression for BCH's formula is~\cite{jacobson1979lie,dynkin2000calculation}
\begin{equation}
	\log(\exp{X}\exp{Y})=\sum_{n=1}^{\infty}\frac{(-1)^{n-1}}{n}\sum_{\substack{r_1+s_1>0 ,\\ \cdots, \\r_n+s_n>0}}\frac{[X^{r_1}Y^{s_1} \cdots X^{r_n}Y^{s_n}]}{\sum_{i=1}^{n}(r_i+s_i)\cdot\prod_{i=1}^n r_i! s_i!}.
    \label{dynkin}
\end{equation}
Here,  the sum is performed over all positive integers $n$, nonnegative combinations of $r_i, s_i$, and
\begin{equation}
{\displaystyle [X^{r_{1}}Y^{s_{1}}\dotsm X^{r_{n}}Y^{s_{n}}]=[\underbrace {X,[X,\dotsm [X} _{r_{1}},[\underbrace {Y,[Y,\dotsm [Y} _{s_{1}},\,\dotsm \,[\underbrace {X,[X,\dotsm [X} _{r_{n}},[\underbrace {Y,[Y,\dotsm Y} _{s_{n}}]]\dotsm ]].}
\end{equation}

For each commutator monomial $C=[X^{r_1}Y^{s_1} \cdots X^{r_n}Y^{s_n}]$, define the \emph{order} $N(C):=\sum_{i=1}^n (r_i+s_i)$. An $N$-th order approximation of $\log(\exp{X}\exp{Y})$ is the summation over all monomial terms with order at most $N$. 

Several difficulties lies ahead. First, each monomial appears multiple times in Dynkin's formula due to different ways of separating one term into $(r,s)$ pairs, leading to inefficiency for the computation. Therefore it would be desirable to come up with a more efficient method for computing the coefficient associated to each term.  Second, there are inherently exponentially many terms need to be taken into account with respect to $N$. Although this cannot be accelerated to polynomial time, we can use several tricks to make it more time and space efficient. Here we focus on the first point. In the next section, we present a more efficient way of calculating the coefficient associated to each monomial term.

\section{Coefficients associated to each monomial}
In this section we focus on computing the coefficient $M(C)$ associated to a given monomial $C$. Note that some monomials in the BCH formula might be linearly dependent so that we can combine the coefficients together; we ignore this issue for now and just focus on the coefficient which arises in the formula itself, i.e.,
$$M(C)=\sum_{n=1}^{\infty}\frac{(-1)^{n-1}}{n}\sum_{r_i, s_i}\frac{1}{\sum_{i=1}^n (r_i+s_i)\cdot\prod_{i=1}^n r_i!s_i!}.$$

where the second summation is over all $n$ pairs $(r_i,s_i)_{i=1}^n$ which gives rise to the monomial $C$. Note that by definition of $N(C)$, $\sum_{i=1}^n (r_i+s_i)=N$ is a fixed number, and
$$N(C)M(C)=\sum_{n=1}^{\infty}\frac{(-1)^{n-1}}{n}M(C,n),$$
where 
$$M(C,n)=\sum_{(r_i,s_i)}\frac{1}{\prod_{i=1}^n r_i!s_i!}.$$

\subsection{Separation into blocks}
Given an $N$th-order monomial $C=[X,[\cdots,[Y,[,\cdots,[\cdots,[X,Y]]]]]]$, we can encode it into an $N$-bit binary string $X\cdots Y\cdots\cdots XY$. Here we identify a monomial with its encoding as a string. $n$ pairs of numbers $(r_i,s_i)_{i=1}^n$ gives rise to $C$ if and only if $C$ is exactly the concatenation $X^{r_1}Y^{s_1}\Vert X^{r_2}Y^{s_2}\Vert\cdots\Vert X^{r_n}Y^{s_n}$. We call such $(r_i,s_i)_{i=1}^n$ a \emph{partition} of the string $C$, and $X^{r_i}Y^{s_i}$ the $i$th substring with respect to the partition $(r_i,s_i)_{i=1}^{n}$. Note that each substring takes the form $X^{r_i}Y^{s_i}$, therefore whenever there is a descending edge $YX$ in the original string $C$, that $Y$ and that $X$ must not lie in the same substring. This enables us to separate the string $C$ to blocks by descending edges, e.g.,
$$C=\underbrace{YY}_{\text{block }1}|\underbrace{XXY}_{\text{block }2}|\underbrace{XY}_{\text{block }3}|\underbrace{XXYY}_{\text{block }4}|\underbrace{X}_{\text{block } 5}.$$
Denote $L(C)$ the number of blocks in $C$ separated by descending edges. Then $$C=X^{u_1}Y^{v_1}\cdots X^{u_{L(C)}}Y^{v_{L(C)}}$$ can be uniquely specified by $L(C)$ pairs of numbers $(u_i,v_i)_{i=1}^{L(C)}$, where $u_i,v_i>0$ except for $u_1$ and $v_{L(C)}$. It is clear that each block contains at least one substrings, yet no substrings can go across blocks. Given that each substring must also be nonempty, we know that the number of substrings separating a nomonimal $C$ is bounded between $L(C)$ and $N(C)$, i.e.,
$$N(C)M(C)=\sum_{n=L(C)}^{N(C)}\frac{(-1)^{n-1}}{n}M(C,n).$$
On the other hand, since no substrings can go across different blocks, it suffices to consider each block separately. Suppose that $n_i$ substrings are allocated to block $i$ with $1\leq i\leq L(C)$, then with fixed sequence $(n_i)_{i=1}^{L(C)}$, how these $n_i$ substrings are allocated inside block $i$ is independent of the allocation inside other blocks, so we can simplify the expression of $M(C,n)$ to be
$$M(C,n)=\sum_{n_1+\cdots +n_{L(C)}=n}\prod_{i=1}^{L(C)}\left(\sum_{(r,s)}\frac{1}{\prod r_j!s_j!}\right),$$
where the inner summation is only over all partitions of block $i$ into $n_i$ substrings. Note that each block $i$ takes the form $X^{u_i}Y^{v_i}$, thus it can be identified by a pair $(u_i,v_i)$. Furthermore, the inner summation only depends on the current block and the number of substrings, so we can denote it as $g(u_i,v_i,n_i)$ and the total summation then becomes
$$M(C,n)=\sum_{n_1+\cdots+n_{L(C)}=n}\prod_{i=1}^{L(C)}g(u_i,v_i,n_i).$$
\subsection{Contribution from individual blocks}
Now let's compute $$g(u_i,v_i, n_i)=\sum_{(r,s)}\frac{1}{\prod r_i!s_i!}.$$ Each term in the summation corresponds to one particular partition $(r_j,s_j)_{j=1}^{n_i}$ of $X^{u_i}Y^{v_i}$ into $n_i$ substrings. Since the block being separated takes the form $X^{u_i}Y^{v_i}$, we know that at most one substring contains both $X$ and $Y$, or equivalently, at most one pair of $(r_j,s_j)$ inside this block has both entries nonzero. Furthermore, given such a partition $(r_j, s_j)$, there exists a partition of the block into $n_i+1$ pieces, which is just refining the substring containing both $X$ and $Y$ to two substrings, one consisting of only $X$s and the other only $Y$s. One can observe that the contribution of coefficients from these two partitions are identical. In the case that both $u_i$ and $v_i$ are nonzero, such a correspondence is one-to-one, meaning that every partition into $n_i$ substrings with only $X^{r_j}$'s and $Y^{s_j}$'s can be mapped to a partition into $n_i-1$ substrings by merging the middle two substrings. We will deal with the case that either $u_i$ or $v_i$ is zero later. Let 
$$h(u_i,v_i,n_i)=\sum_{(r,s)}\frac{1}{\prod r_j!s_j!}$$
be the summation over all partitions without substrings containing both $X$ and $Y$, then 
$$g(u_i,v_i,n_i)=h(u_i,v_i,n_i)+h(u_i,v_i,n_i+1).$$
Since no string contains both $X$ and $Y$, we can enumerate over the number of substrings partitioning $X^{u_i}$ and $Y^{v_i}$, then
$$h(u_i,v_i,n_i)=\sum_{1<n_x<n_i}\sum_{\substack{r_1,\cdots, r_{n_x}>0\\\sum r_j=u_i}}\sum_{\substack{s_{n_x+1},\cdots, s_{n_i}>0\\\sum s_j=v_i}}\frac{1}{\prod_{j=1}^{n_x} r_j!\prod_{j=n_x+1}^{n_i}s_j!}.$$
Denote $f(u,n)=\sum_{\substack{r_1,\cdots, r_n>0,\\\sum_j{r_j}=u}}\frac{1}{\prod_{j=1}^n r_j!}$, then we have
$$h(u_i,v_i,n_i)=\sum_{1<n_x<n_i}f(u_i,n_x)*f(v_i,n_i-n_x).$$

Rewrite $f(u,n)$ as
$$f(u,n)=\frac{1}{u!}\sum_{\substack{r_1,\cdots, r_n>0,\\\sum_j{r_j}=u}}\frac{u!}{\prod_{j=1}^n r_j!}=\frac{1}{u!}\sum_{\substack{r_1,\cdots, r_n>0,\\\sum_j{r_j}=u}}\binom{u}{r_1,\cdots, r_n}.$$
By multinomial theorem, we know that $\sum_{\substack{r_1,\cdots, r_n\geq0,\\\sum_j{r_j}=u}}\binom{u}{r_1,\cdots, r_n}=n^u$. This summation is almost the term we want except that it has extra terms where some of the $r_j$s are zero. By inclusion-exclusion principle, we have
$$u!f(u,n)=\sum_{S\subseteq [n]}(-1)^{|S|}\sum_{\substack{r_j\geq 0,j\notin S,\\r_j=0,j\in S,\\ \sum_j{r_j}=u}}\binom{u}{r_1,\cdots, r_n}=\sum_{S\subseteq [n]}(-1)^{|S|}(n-|S|)^u=\sum_{z=0}^n(-1)^z\binom{n}{z}(n-z)^u.$$
We can express $f(u,n)$ more concisely in terms of finite difference as
$$f(u,n)=\frac{1}{u!}\Delta^n_x x^u|_{x=0}.$$

The case where either $u_i$ or $v_i$ is zero can be similarly calculated; we have
$$g(u_i,0,n_i)=g(0,u_i,n_i)=f(u_i,n_i).$$

\subsection{Computing the overall coefficient}
Given all blocks, we are now ready to compute the coefficient given a monomial $C$. We first divide $C$ into blocks $(u_i,v_i)_{i=1}^{L(C)}$. Then
$$M(C)=\underbrace{\frac{1}{N(C)\cdot\prod_{i=1}^{L(C)}u_i!v_i!}}_{\text{overall constant}}\cdot \underbrace{\sum_{n=L(C)}^{N(C)}}_{S_1}\frac{(-1)^{n+1}}{n}\underbrace{\sum_{\substack{n_1,\cdots, n_{L(C)}>0\\ \sum_{i=1}^{L(C)}n_i=n}}}_{S_2}\underbrace{\prod_{i=1}^{L(C)}}_{P_1}\underbrace{g'(u_i,v_i,n_i)}_{T},$$
$$g'(u_i,v_i,n_i)=\begin{cases}\sum_{1<n_x<n_i}f'(u_i,n_x)*f'(v_i,n_i-n_x)+\sum_{1<n_x<n_i+1}f'(u_i,n_x)*f'(v_i,n_i-n_x+1), u_i,v_i>0,\\ f'(u_i,n_i), v_i=0,\\ f'(v_i,n_i), u_i=0,\end{cases}$$
$$f'(u,n)=\Delta^n_x x^u|_{x=0}=\sum_{z=0}^n(-1)^z\binom{n}{z}(n-z)^u.$$
\subsection{*Complexity analysis}
The main complexity of computing such a coefficient comes from the nested function calls $S_1$, $S_2$, $P_1$ and the subroutine $T$ computing $g'(u_i,v_i,n_i)$, since the overall constant can be computed only once so it does not contribute much to the complexity. $P_1$ has $L(C)$ factors; $T$ can either be computed from scratch in $O(N^3)$ time, or be computed from preprocessed table storing $f(u,n)$'s in $O(N)$ time with $O(N^2)$ extra memory, or in $O(1)$ time from preprocessed table storing $g(u,v,n)$, with $O(N^3)$ extra memory. Putting $S_2$ and $S_1$ together, we are essentially summing up over all possible numbers of substrings inside each block. For a block with length $N_i$, it can be partitioned into $1$ to $N_i$ substrings, and the number of substrings inside this block is independent over the numbers of substrings inside other blocks. Altogether, there are $\prod_{i=1}^{L(C)}N_i\leq e^{N/e}$ summands to take into consideration. Putting everything together, the complexity of computing $M(C)$ for a monomial $C$ can be reduced to
$O(e^{N/e})$ with $O(N^3)$ extra memory. Since the complexity for computing the coefficient dominates the cost for computing the commutator itself, enumerating over all strings with length up to $N$, the total running time would be $O(2^N\cdot e^{N/e})=O(2^{1.53N})$. A more careful analysis might give a tighter bound (numerical evidence shows that the running time scales as $O(2^{1.47N})$), but for now it is not our main focus. One can see that it is a big improvement with respect to naively enumerating over all possible partitions for each monomial $C$, which would take $\Omega(2^{2N})$ time.

\bibliographystyle{plain}
\bibliography{sample}

\end{document}